# On context-free languages of scattered words[*]

Z. Ésik
Dept. of Computer Science
University of Szeged
Hungary

S. Okawa
School of Computer Science and Engineering
University of Aizu
Japan

November 7, 2018

**Abstract**

It is known that if a Büchi context-free language (BCFL) consists of scattered words, then there is an integer $n$, depending only on the language, such that the Hausdorff rank of each word in the language is bounded by $n$. Every BCFL is a Müller context-free language (MCFL). In the first part of the paper, we prove that an MCFL of scattered words is a BCFL iff the rank of every word in the language is bounded by an integer depending only on the language.

Then we establish operational characterizations of the BCFLs of well-ordered and scattered words. We prove that a language is a BCFL consisting of well-ordered words iff it can be generated from the singleton languages containing the letters of the alphabet by substitution into ordinary context-free languages and the $\omega$-power operation. We also establish a corresponding result for BCFLs of scattered words and define expressions denoting BCFLs of well-ordered and scattered words. In the final part of the paper we give some applications.

## 1 Introduction

A word over an alphabet $A$ is an isomorphism type of a labeled linear order. In this paper, in addition to finite and $\omega$-words, we also consider words whose underlying linear order is any countable linear ordering, cf. [34]. Countable words and in particular regular words were first investigated in [16], where they were called "arrangements". Regular words were later studied in [4, 6, 7, 28, 35] and more recently in [31]. Context-free words were introduced in [8] and their underlying linear orderings were investigated in [9, 10, 18, 19, 20].

[*]The first author was partially supported by the project TÁMOP-4.2.1/B-09/1/KONV-2010-0005 "Creating the Center of Excellence at the University of Szeged", supported by the European Union and co-financed by the European Regional Fund, and by the grant no. K 75249 from the National Foundation of Hungary for Scientific Research.

Finite automata on $\omega$-words have by now a vast literature, see [33] for a comprehensive treatment. Also, finite automata acting on well-ordered words longer than $\omega$ have been investigated by many authors, a small sampling is [1, 13, 14, 37, 38]. In the last decade, the theory of automata on well-ordered words has been extended to automata on all countable words, including scattered and dense words. In [2, 3, 12], both operational and logical characterizations of the class of languages of countable words recognized by finite automata were obtained.

Context free grammars generating $\omega$-words were introduced in [15] and subsequently studied in [11, 32]. Context-free grammars generating arbitrary countable words were defined in [22, 23]. Actually, two types of grammars were defined, context-free grammars with Büchi acceptance condition (BCFG), and context-free grammars with Müller acceptance condition (MCFG). These grammars generate the Büchi and the Müller context-free languages of countable words, abbreviated as BCFLs and MCFLs. It is clear from the definitions in [22, 23] that every BCFL is an MCFL. On the other hand, there exist MCFLs of even well-ordered words that are not BCFLs, for example the set of all countable well-ordered words over some alphabet. This is due to the fact that the order-type of every word in a BCFL of well-ordered words is bounded by the ordinal $\omega^n$, for some integer $n$ depending on the language, cf. [22]. More generally, it was shown in [22] that for every BCFL $L$ of scattered words there is an integer $n$ such that the Hausdorff rank of every word in $L$ is bounded by $n$. On the other hand, regarding MCFLs $L$ of scattered words, two cases arise, cf. [23]. Either there exists an integer $n$ such that the rank of every word in $L$ is bounded by $n$, or for every countable ordinal $\alpha$ there is a word in $L$ whose Hausdorff rank exceeds $\alpha$. It is then natural to ask whether every MCFL of scattered words of the first type is a BCFL. In this paper, we answer this question: all such MCFLs are in fact BCFLs. Thus, the BCFLs of scattered words are exactly the "bounded" MCFLs of scattered words.

Then we establish operational characterizations of the BCFLs of well-ordered and scattered words. We prove that a language is a BCFL consisting of well-ordered words iff it can be generated from the singleton languages containing the letters of the alphabet by substitution into ordinary context-free languages and the $\omega$-power operation. We also establish a corresponding result for BCFLs of scattered words and define expressions denoting BCFLs of well-ordered and scattered words. In the final part of the paper, we give some applications of the main results.

## 2 Basic notions

### 2.1 Linear orderings

A linear ordering $(I, <)$ consists of a set $I$ and a strict linear order relation $<$ on $I$. When the set $I$ is finite or countable, we call $(I, <)$ finite or countable as well. *In the rest of the paper, by a linear ordering we will always mean a countable ordering.* A good reference for linear orderings is [34].

A morphism of linear orderings $(I, <) \to (J, <')$ is a function $h : I \to J$ that preserves the order relation, so that for all $x, y \in I$, if $x < y$ then $h(x) <' h(y)$. Since every morphism is an injective function, we sometimes call a morphism an *order embedding*, or just an embedding. If $I \subseteq J$ and the inclusion $I \hookrightarrow J$ is an order embedding, then we say that $I$ is *sub-ordering* of $J$. When $(I, <)$ is a subordering of $(J, <')$, the relation $<$ is the restriction of $<'$ onto $I$. An *isomorphism* is a bijective morphism. Isomorphic linear orderings have the same *order type*. The order type of a well-ordering is a (countable) ordinal. We identify the finite ordinals with the nonnegative integers.

Some examples of linear orderings are the usual orderings of the nonnegative integers $(\mathbb{N}_+, <)$, the ordering of the negative integers $(\mathbb{N}_-, <)$, and the ordering $(\mathbb{Q}, <)$ of the rationals. Their respective order types are denoted $\omega$, $-\omega$ and $\eta$.

Let $(I, <)$ be a linear ordering. We say that $(I, <)$ is a *well-ordering* if each nonempty subset of $I$ has a least element. This condition is equivalent to requiring that $(I, <)$ has no sub-ordering of order type $-\omega$. Moreover, we say that $(I, <)$ is *dense* if it has at least two elements and for all $x, y \in I$ with $x < y$ there is some $z \in I$ with $x < z < y$. Finally, we say that $(I, <)$ is *scattered* if it has no dense sub-ordering, and *quasi-dense* if it is not scattered. It is well-known that every sub-ordering of a well-ordering is well-ordered, and every sub-ordering of a scattered ordering is scattered. Moreover, up to isomorphism there are four (countable) dense linear orderings, the ordering of the rationals possibly endowed with a least or a greatest element, or both. The respective order types of these linear orderings are $\eta$, $1 + \eta$, $\eta + 1$ and $1 + \eta + 1$. (See below for the sum operation on order types.)

When $(I_1, <_1)$ and $(I_2, <_2)$ are linear orderings, their *sum* $(I_1, <_1) + (I_2, <_2)$ is the linear ordering $(I, <)$, where $I = (I_1 \times \{1\}) \cup (I_2 \times \{2\})$, moreover, for all $(x, i), (y, j) \in I$, $(x, i) < (y, j)$ iff $i = 1$ and $j = 2$, or $i = j$ and $x <_i y$. The sum operation may be generalized. Suppose that $(J, <)$ is a linear ordering, and for each $j \in J$, $(I_j, <_j)$ is a linear ordering. Then the *generalized sum* $\sum_{j \in J}(I_j, <_j)$ is the disjoint union

$$\biguplus_{j \in J} I_j = \{(x, j) : j \in J, \ x \in I_j\}$$

equipped with the order relation $(x, j) < (y, k)$ iff $j < k$, or $j = k$ and $x <_j y$. We call a generalized sum a well-ordered, a scattered, or a dense sum, when $(J, <)$ has the appropriate property. It is known that every well-ordered sum of well-orderings is a well-ordering, and similarly, every scattered sum of scattered orderings is scattered and every dense sum of dense orderings is dense. When each $(I_j, <_j)$ is the linear ordering $(I, <')$, then the generalized sum $\sum_{j \in J}(I_j, <_j)$ is called the *product* of $(I, <')$ and $(J, <)$, denoted $(I, <') \times (J, <)$. When $(I, <)$ and $(J, <)$ are both well-ordered, scattered or dense, then so is their sum or product. Since the above operations preserve isomorphism, they can be extended to order types.

Hausdorff classified scattered linear orderings into an infinite hierarchy. Following [30], we present a variant of this hierarchy. Let $VD_0$ be the collection of all finite linear orderings, and for a countable ordinal $\alpha > 0$, let $VD_\alpha$ be the collection of all finite sums of linear orderings of the sort

$$\sum_{n \in \mathbb{N}_+} (I_n, <_n) \quad \text{or} \quad \sum_{n \in \mathbb{N}_-} (I_n, <_n),$$

where each $(I_n, <_n)$ is in $VD_{\beta_n}$ for some $\beta_n < \alpha$. By Hausdorff's theorem [34], a linear ordering $(I, <)$ is scattered iff it belongs to $VD_\alpha$ for some (countable) ordinal $\alpha$. The least such ordinal is called the *rank* of $I$, denoted r($I$). Hausdorff also proved that every linear ordering is either scattered, or a dense sum of scattered linear orderings.

A useful fact is that a well-ordering has rank $\alpha$ iff its order type $\gamma$ satisfies $\omega^\alpha \leq \gamma < \omega^{\alpha+1}$, so that its *Cantor normal form* is

$$\omega^\alpha \times n_0 + \omega^{\alpha_1} \times n_1 + \ldots + \omega^{\alpha_k} \times n_k,$$

where $k \geq 0$, $\alpha > \alpha_1 > \ldots > \alpha_k$ and $n_0, \ldots, n_k$ are positive integers.

## 2.2 Words and languages

A *word* (or *arrangement* [16]) $u$ over a possibly infinite alphabet $A$ is a linear ordering $I = (I, <)$ labeled in $A$. Thus a word $u$ is of the form $(I, <, \lambda)$, where $\lambda : I \to A$. A *morphism* between words preserves the order relation and the labeling. An *isomorphism* is a bijective morphism. We usually identify isomorphic words. The *order type* of a word is the order type of its underlying linear order.

Examples of words include the finite words whose underlying linear order is finite, including the *empty word* $\epsilon$ whose order type is 0, the one-letter words $a^\omega$ and $a^{-\omega}$, labeled $a$, whose underlying linear orders are the orderings of the nonegative and the negative integers, and the one-letter word $a^\eta$ whose underlying linear order is the ordering of the rationals.

We call a word *well-ordered, scattered, dense*, or *quasi-dense* if its underlying linear order has the appropriate property. The *rank* r($u$) of a scattered word $u$ is the rank of its underlying linear ordering. For example, $a^\omega$ is well-ordered, $a^{-\omega}$ is scattered but not well-ordered, and $a^\eta$ is dense. Also, $r(a^{-\omega}) = r(a^\omega) = 1$. More generally, when $\alpha$ is a countable ordinal, $a^\alpha$ is the word whose underlying linear order is a well-order of order type $\alpha$, with each point labeled $a$. The word $a^\omega a^\eta$ obtained by "concatenating" $a^\omega$ and $a^\eta$ is quasi-dense, but not dense. (A formal definition of concatenation is given below).

Let $A^\sharp$ denote the set of all words over $A$. As usual, we denote by $A^*$ and $A^\omega$ the sets of all finite and all $\omega$-words over $A$, whose order type is $\omega$. We define $A^{\leq \omega} = A^* \cup A^\omega$.

A *language* over $A$ is any subset of $A^\sharp$. In particular, every subset of $A^*$ is a language (of finite words). Languages over $A$ are equipped with the usual set theoretic operations. We now define the operation of *substitution*.

Suppose that $L \subseteq A^\sharp$, and for each $a \in A$, $L_a \subseteq B^\sharp$. Then the language

$$L[a \mapsto L_a]_{a \in A}, \quad \text{or simply} \quad L[a \mapsto L_a]$$

is the language over $B$ consisting of all words obtained from the words $u$ in $L$ by replacing each occurrence of a letter $a \in A$ in $u$ by a word $v \in L_a$. Different occurrences of the same latter may be replaced by different words. Formally, suppose that $u = (I, <, \lambda) \in L$, and for each $i \in I$, let $v_i = (I_i, <_i, \lambda_i)$ be a word in $L_{\lambda(i)}$. Then we construct the word $u'$ whose underlying linear order is the ordered sum $I = \sum_{i \in I}(I_i, <_i)$ which is equipped with the labeling function
$$\lambda'((x, i)) = \lambda_i(x)$$
for all $i \in I$ and $x \in I_i$. The language $L[a \mapsto L_a]_{a \in A}$ consists of all such words $u'$. If $L$ and the $L_a$ contain only well-ordered, scattered or only dense words, then the same holds for $L[a \mapsto L_a]_{a \in A}$. Below we will often follow the convention of writing $L[a \mapsto L_a]_{a \in A_0}$ where $a$ ranges over a subset $A_0$ of $A$ to denote the substitution where each letter $a \in A_0$ is replaced by $L_a$ and each letter not in $A_0$ remains unchanged, i.e., is replaced by $\{a\}$.

When $L$ and each $L_a$ consists of a single word, say $L = \{u\}$ and $L_a = \{v_a\}$, then $L[a \mapsto L_a]$ is also a singleton, and we denote its single element by $u[a \mapsto v_a]$. If $u$ and each $v_a$ is well-ordered (resp. scattered, dense), then so is $u[a \mapsto v_a]$.

Using the generic operation of substitution, we now define further operations on languages. Let $x_0, x_1, \ldots$ be letters. Suppose that $L, L_1, L_2 \subseteq A^\sharp$. Then we define
$$\begin{aligned} L_1 L_2 &= \{x_1 x_2\}[x_i \mapsto L_i] \\ L^\omega &= \{x_0 x_1 \ldots\}[x_i \mapsto L] = \{u_0 u_1 \ldots : u_i \in L\} \\ L^{-\omega} &= \{\ldots x_1 x_0\}[x_i \mapsto L] = \{\ldots u_1 u_0 : u_i \in L\} \end{aligned}$$

When $L = \{u\}$, $L_1 = \{u_1\}$ and $L_2 = \{u_2\}$ are singleton languages, we obtain the word operations of concatenation $u_1 u_2$ and the unary $\omega$-power and $(-\omega)$-power operations $u^\omega = uu\ldots$ and $u^{-\omega} = \ldots uu$.

## 2.3 Context-free languages

When $G = (N, A, R, S)$ is an ordinary *context-free grammar* (CFG), where $N$ is the set of *nonterminals*, $A$ is the finite alphabet of *terminals*, $R$ is the set of *rules* and $S \in N$ is the *start symbol*, we may consider possibly infinite *derivation trees* over $G$. Such a tree is a finitely branching rooted, ordered directed tree labeled in $N \cup A \cup \{\epsilon\}$ such that whenever a vertex $x$ is labeled $X \in N$ and has $n$ successors, ordered as $x_1, \ldots, x_n$ and labeled $X_1, \ldots, X_n \in N \cup A$, then $X \to X_1 \ldots X_n \in R$. When $n = 1$, it is also allowed that $x_1$ is labeled $\epsilon$ and then $X \to \epsilon \in R$. The label of the root is called the *root symbol*. A vertex with no successors is called a *leaf*. In particular, every vertex labeled in $A \cup \{\epsilon\}$ is a leaf. The leaves of a derivation tree $t$ form a linearly ordered set with respect to the usual left-to-right ordering, and considering only the leaves labeled in $N \cup A$, we obtain a word in $(N \cup A)^\sharp$. This word is called the *frontier* of $t$. When the root symbol of a *finite* derivation tree is $X \in N \cup A$ and its frontier is $p \in (N \cup A)^*$, then we write $X \Rightarrow^* p$. As usual, we extend $\Rightarrow^*$ to a binary relation over $(N \cup A)^*$. The *context-free language* (CFL) generated by $G$ is $L(G) = \{u \in A^* : S \Rightarrow^* u\}$.

Suppose that $A$ is a finite alphabet. A *Büchi context-free grammar* (BCFG) over $A$ is a CFG $(N, A, R, S)$ equipped with a designated subset $N_\infty$ of the nonterminals $N$. When $G = (N, A, R, S, N_\infty)$ is a BCFG, call a derivation tree *proper* if along each infinite path (originating in the root) there are infinitely many vertices labeled in $N_\infty$. When the root of such a tree is labeled $X$ and its frontier is the word $p \in (N \cup A)^\sharp$, we write $X \Rightarrow^\infty p$ (or $X \Rightarrow^* p$ when the tree is finite) and say that $p$ is *derivable* from $X$. The *language $L(G)$ generated by $G = (N, A, R, S, N_\infty)$* is the set of all words $u \in A^\sharp$ such that $S \Rightarrow^\infty u$. We say that $L \subseteq A^\sharp$ is a *Büchi context-free language* (BCFL), if $L = L(G)$ for some BCFG $G$.

We also define *Müller context-free grammars* (MCFG) that are CFGs $(N, A, R, S)$ equipped with a set $\mathcal{F} \subseteq P_+(N)$ of nonempty subsets of $N$. We say that a derivation tree is *proper* if for each infinite path, the set of nonterminals that label an infinite number of vertices along the path belongs to $\mathcal{F}$. When $X$ is the root label of a proper derivation tree having frontier $p$, then we write $X \Rightarrow^\infty p$, or $X \Rightarrow^* p$ when the tree is finite. The language $L(G)$ generated by such a grammar $G = (N, A, R, S, \mathcal{F})$ consists of those words $u \in A^\sharp$ such that there is a proper derivation tree $t$ whose root is labeled $S$ having frontier $u$, in notation, $S \Rightarrow^\infty u$. We say that a language $L \subseteq A^\sharp$ is a *Müller context-free language* (MCFL) if $L = L(G)$ for some MCFG $G$. We say that two BCFGs or MCFGs are *equivalent* if they generate the same language.

It is clear that every BCFL is an MCFL. It is not difficult to see that a language $L \subseteq A^*$ is a BCFL iff it is an MCFL iff it is an ordinary context-free language (CFL), cf. [22, 23]. On the other hand, there exists an MCFL that is not a BCFL, for example the set of all well-ordered words over a one-letter alphabet, cf. [22].

BCFLs and MCFLs are closely related to *Büchi* and *Müller tree automata* [33, 36], since a language is a BCFL (MCFL, resp.) iff it is the frontier language of a tree language recognized by a Büchi tree automaton (Müller tree automaton).

We say that a BCFG or an MCFG *has no useless symbols* if either it has a single nonterminal, the start symbol $S$, and no rules, or for each nonterminal $X$ there are finite words $p, q$ and a possibly infinite terminal word $u$ with $S \Rightarrow^* pXq$ and $X \Rightarrow^\infty u$. It then follows that there exist also terminal words $v, w \in A^\sharp$ with $S \Rightarrow^\infty vXw$.

It is known that for each BCFG (MCFG, resp.) there is an equivalent BCFG (MCFG, resp.) having no useless nonterminals.

## 3 Linear context-free languages

In this section, we define linear BCFLs and MCFLs and prove their equivalence. We will later use these linear languages as building blocks to construct more general BCFLs and MCFLs of scattered words.

Recall that a CFG $G = (N, A, R, S)$ is called *linear* if the right-hand side of each rule in $R$ contains at most one occurrence of a nonterminal. A *linear language* in $A^*$ is the language generated by a linear grammar. We call a BCFG (MCFG, respectively) linear if its underlying context free grammar is linear. A linear BCFL (MCFL, respectively) is a BCFL (MCFL) that is generated by a linear BCFG (MCFG). Since every BCFL is an MCFL, every linear BCFL is a linear MCFL.

Note that when $G$ is a linear, then every derivation tree has a single maximal path that contains all the nonterminal labeled vertices. We call this path the *principal path* of the derivation tree. Every vertex that does not belong to the principal path is a leaf labeled in $A \cup \{\epsilon\}$. It follows from this fact that the order type of each word of a linear BCFL or MCFL is either a finite ordinal $n$, or of the form $\omega + n$, $n + (-\omega)$ or $\omega + (-\omega)$. Thus, every word of a linear BCFL or MCFL is scattered of rank at most 1.

Linear BCFLs and MCFLs are closely related to *Büchi automata* and *Müller automata*, cf. [33]. By a Büchi-automaton we mean a system $\mathcal{A} = (Q, A, \delta, q_0, F, Q_\infty)$, where $Q$ is the finite nonempty set of *states*, $A$ is the finite *input alphabet*, $\delta \subseteq Q \times A \times Q$ is the *transition relation*, $q_0 \in Q$ is the *initial state*, $F \subseteq Q$ is the set of *final states* and $Q_\infty$ is a *designated subset of* $Q$. A *run* of $\mathcal{A}$ on a word $u \in A^{\leq \omega}$ is defined as usual. A *run on a finite word is successful* if it starts in the initial state and ends in a state in $F$. A *run on an $\omega$-word is successful* if it visits at least one state in $Q_\infty$ infinitely often. The *language accepted by* $\mathcal{A}$ consists of all words $u \in A^{\leq \omega}$ such that $\mathcal{A}$ has a successful run on $u$. A Müller automaton $\mathcal{A} = (Q, A, \delta, q_0, F, \mathcal{F})$ is defined similarly, but instead of a subset of $Q$, the last component $\mathcal{F}$ is a *designated subset of* $P_+(Q)$. An infinite run is called successful if it starts in the initial state and the set of states visited infinitely often belongs to $\mathcal{F}$. The *language accepted by a Müller automaton* $\mathcal{A}$ is the set of all words in $u \in A^{\leq \omega}$ on which $\mathcal{A}$ has a successful run. It is well-known that a language is accepted by a Büchi automaton iff it is accepted by a Müller automaton. The notion of Büchi automata and Müller automata may be generalized without altering the computation power by allowing a finite number of transitions of the form $(q, u, q')$ where $q, q' \in Q$ and $u \in A^*$.

LEMMA 3.1 *Every linear MCFL is a BCFL.*

*Proof.* Suppose that $G = (N, A, R, S, \mathcal{F})$ is a linear MCFG. Let us construct the following Müller automaton $\mathcal{A}$:

- The set of states is $N \cup \{Z_0\}$, where $Z_0$ is a new symbol.
- The set of terminals is $R$.
- The set of transitions consists of the triples $(X, r, Y)$ such that $r$ is a rule of the form $X \to uYv$ for some $u, v \in A^*$, together with all transitions $(X, r, Z_0)$ such that $r$ is a rule of the form $X \to u$ with $u \in A^*$.
- The initial state is $S$.
- The set of final states is $\{Z_0\}$.
- The designated subsets of the state set are those in $\mathcal{F}$.

Clearly, this Müller automaton accepts all words in $R^{\leq \omega}$ which arise as the sequence of rules applied along the principal path of some derivation tree rooted $S$ whose frontier is a terminal word.

This Müller automaton has an equivalent Büchi automaton, say $\mathcal{B} = (Q, R, \delta, q_0, F, Q_\infty)$, where $F$ is the set of final states and $Q_\infty$ is the set of designated states. Then let us construct the BCFG $G' = (Q, A, R', q_0, Q_\infty)$, where $R'$ consists of all rules $q \to uq'v$ such that for some $r \in R$, $(q, r, q') \in \delta$ and $r$ is a rule of the form $X \to uYv$ for some $X, Y \in N$, together with all rules of the form $q \to u$ such that for some $r \in R$ and $q' \in F$, $(q, r, q') \in \delta$, moreover, $r$ is of the form $X \to u$ for some $X$. Since $\mathcal{B}$ is equivalent to $\mathcal{A}$, it follows from the description of the language accepted by $\mathcal{A}$ that $G$ is equivalent to $G'$. $\square$

An operational characterization of linear BCFLs was given in [21]. In order to recall this result, we extend the $\omega$-power operation to sets of pairs of words.

When $A$ is a finite alphabet, we may consider ordered pairs $(u, v) \in A^\sharp \times A^\sharp$ that form a monoid with respect to the *product operation*

$$(u, v)(u', v') = (uu', v'v)$$

with the pair $(\epsilon, \epsilon)$ acting as identity. Then we may consider the power set of this monoid, $P(A^\sharp \times A^\sharp)$, and equip this set with the operations of set union and *complex product*:

$$U \cdot V = \{(u, v)(u', v') : (u, u') \in U, \ (v, v') \in V\} = \{(uu', v'v) : (u, u') \in U, \ (v, v') \in V\}.$$

With these operations and the constants $\emptyset$ and $\{(\epsilon, \epsilon)\}$, $P(A^\sharp \times A^\sharp)$ is an *idempotent semiring* [26]. We may also define a *star operation* by

$$U^* = \bigcup_{n \geq 0} U^n.$$

The set $P(A^\sharp)$ of all subsets of $A^\sharp$ is a commutative idempotent monoid with respect to the operation of set union and the constant $\emptyset$. We define an action of the semiring $P(A^\sharp \times A^\sharp)$ on $P(A^\sharp)$ by

$$U \circ L = \{uwv : (u, v) \in U, \ w \in L\}.$$

Moreover, we define an $\omega$-power operation $P(A^\sharp \times A^\sharp) \to P(A^\sharp)$,

$$U \mapsto \{(u_0 u_1 \ldots) \cdot (\ldots v_1 v_0) : (u_i, v_i) \in U\}.$$

Note that the $\omega$-power operation defined earlier on languages $L \subseteq A^\sharp$ can now be expressed as $L^\omega = (L \times \{\epsilon\})^\omega$, and the $(-\omega)$-power operation by $L^{-\omega} = (\{\epsilon\} \times L)^\omega$. (The semiring-semimodule pair $((P(A^\sharp \times A^\sharp), P(A^\sharp))$ equipped with the star and $\omega$-power operations is in fact an *iteration semiring-semimodule pair*, cf. [5, 24].)

Using algebraic tools, the following Kleene theorem for linear BCFLs was proved in [21] as a special case of a more general theorem. It is also possible to derive this result from the classical Kleene theorem for regular $\omega$-languages, cf. [33].

THEOREM 3.2 *A language $L \subseteq A^\sharp$ is a linear BCFL iff it is a finite union of languages of the sort*

$$U \circ V^\omega,$$

where $U$ and $V$ can be generated from the finite subsets of $A^* \times A^*$ by the operations $\cup$, $\cdot$ and $^*$.

It follows from this result also that the Hausdorff rank of every word in a linear BCFL is at most 1.

COROLLARY 3.3 *A language $L \subseteq A^\sharp$ is a linear BCFL iff it is a union of an ordinary linear CFL $L_0 \subseteq A^*$ with a finite union of languages of the sort*

$$U \circ V^\omega$$

*where $U$ and $V$ can be generated from the finite subsets of $A^* \times A^*$ by the operations $\cup$, $\cdot$ and $^*$, moreover, $U$ is nonempty and $V$ contains at least one pair one of whose components is not $\epsilon$.*

COROLLARY 3.4 *A language $L \subseteq A^\sharp$ of well-ordered words is a linear BCFL iff it is a union of an ordinary linear CFL $L_0 \subseteq A^*$ with a finite union of languages of the sort*

$$K_0 K_1^\omega$$

*where $K_0, K_1 \subseteq A^*$ are ordinary nonempty regular languages and $K_1$ contains at least one nonempty word.*

*Proof.* This follows from the previous corollary by noting that when $U$ and $V$ are not empty, then $U \circ V$ contains only well-ordered words iff the second component of each pair in $U \cup V$ is $\epsilon$. □

Note that the order type of every word of a BCFL of well-ordered words is at most $\omega$.

COROLLARY 3.5 *A language $L \subseteq A^\sharp$ is a linear BCFL iff it is a language $L \subseteq A^{\leq \omega}$ that can be accepted by a Büchi automaton (that is regular).*

## 4 Context-free languages of scattered words of bounded rank

Call a language $L$ of scattered words *bounded*[1] if the rank of the words of $L$ is bounded by an integer. It was proved in [22] that every BCFL of scattered words is bounded. In this section our aim is to prove that when $L$ is a bounded language of scattered words, then $L$ is an MCFL iff $L$ is a BCFL. We will derive this result from the fact, established below, that every language generated by a "non-reproductive MCFG" is a BCFL.

When $G = (N, A, R, S)$ is a CFG, the graph $\Gamma_G$ has $N$ as its vertex set and edges $X \to Y$ if there is a rule of the form $X \to pYq$. We say that a nonterminal $Y$ is

---
[1]This notion has nothing to do with the classical notion of a bounded language $L \subseteq A^*$.

*accessible* from $X$ if there is a path from $X$ to $Y$ in $\Gamma_G$. A subset $N' \subseteq N$ is *strongly connected* if for all $X, Y \in N'$, $Y$ is accessible from $X$. A *strong component* is a maximal strongly connected subset. The *height* of a nonterminal $X$ is the length $n$ of the longest sequence $Y_0, Y_1, \ldots, Y_n$ of nonterminals belonging to different strong components such that $Y_n = X$ and $Y_{i-1}$ is accessible from $Y_i$ for each $1 \leq i \leq n$. The above notions all extend to BCFGs and MCFGs.

We recall from [23] that a nonterminal $X$ of an MCFG $G = (N, A, R, S, \mathcal{F})$ is *reproductive* if there is a word $p \in (N \cup A)^\sharp$ containing an infinite number of occurrences of $X$ with $X \Rightarrow^\infty p$. We call $G$ *non-reproductive* if it has no reproductive nonterminal.

THEOREM 4.1 *Suppose that $G = (N, A, R, S, \mathcal{F})$ is a non-reproductive MCFG. Then there is a BCFG equivalent to $G$.*

*Proof.* Suppose that $L = L(G)$ for a non-reproductive MCFG $G = (N, A, R, S, \mathcal{F})$. We may suppose that $L(G)$ is nonempty, since in the opposite case our claim is obvious. Since for each infinite path of a derivation tree, the nonterminals visited infinitely often form a strongly connected set, without loss of generality we may assume that each $F \in \mathcal{F}$ is itself strongly connected and thus included in some strong component.

Let $X$ be a nonterminal of height $n$. We show by induction on $n$ that the language $L(X) = \{u \in A^\sharp : X \Rightarrow^\infty u\}$ generated from $X$ is a BCFL. Suppose that we have proved this claim for all nonterminals of height $< n$. Let $C$ denote the strong component of $X$. If $Y \to u$ is in $R$ with $Y \in F$, for some $F \in \mathcal{F}$, $F \subseteq C$, then $u$ contains at most one occurrence of a nonterminal in $F$, and if it does, then no other nonterminal in $C$ occurs in $u$, see also [29].

Indeed, suppose to the contrary that $Y \to pZq \in R$, where $Y, Z \in F$ for some $F \in \mathcal{F}$ with $F \subseteq C$, and suppose that $pq$ contains an occurrence of a nonterminal in $C$. Since $F$ is strongly connected, there is a sequence of rules

$$Z \to p_1 Z_1 q_1, \ldots, Z_{m-1} \to p_m Y q_m$$

with $F = \{Y, Z, Z_1, \ldots, Z_{m-1}\}$. Let $p' = p_1 \ldots p_m$, $q' = q_m \ldots q_1$. Then $Y \Rightarrow^\infty (pp')^\omega (q'q)^{-\omega}$. By assumption, $p$ or $q$ contains a nonterminal in $C$. By symmetry, we may assume that $p$ contains such a nonterminal. Since $C$ is strongly connected, there is a finite derivation $p \Rightarrow^* rYs$ for some words $r, s$. Thus we have $Y \Rightarrow^\infty (rYsp')^\omega (q'q)^{-\omega}$, contradicting the assumption that $G$ is non-reproductive.

Let $N'$ denote the set of all nonterminals not in $C$ accessible from $X$. For each $F \in \mathcal{F}$, $F \subseteq C$ and $Y \in F$, let us consider the MCFG $G_{F,Y}$ whose set of nonterminals is $F$ and whose terminals are the letters in $N' \cup A$. The start symbol is $Y$ and the only designated subset of $F$ is $F$ itself. The rules are those rules of $R$ whose left side is in $F$ and whose right side contains a nonterminal in $F$. Note that $G_{F,Y}$ is linear. Let $L_{F,Y} \subseteq (N' \cup A)^\sharp$ denote the language generated by the MCFG $G_{F,Y}$.

Moreover, let $G_X$ denote the ordinary CFG whose set of nonterminals is $C$, set of terminals is $N' \cup A \cup C'$, where $C' = \{(F, Y) : Y \in F \subseteq C, \ F \in \mathcal{F}\}$, and whose start symbol is $X$. The rules of $G_X$ are all rules in $R$ whose left side belongs to $C$, together with all rules $Y \to (F, Y)$ with $Y \in F \subseteq C$, $F \in \mathcal{F}$. Let $L_X \subseteq (N' \cup A \cup C')^*$ denote the ordinary context-free language generated by $G_X$.

Now the language $L(X) \subseteq A^\sharp$ can be constructed as follows. First, for each $(F, Y)$ as above, let us substitute words of $L(Z) \subseteq A^\sharp$ for each occurrence of each nonterminal $Z \in N'$ in the words of $L_{F,Y}$ to obtain the language

$$L'_{F,Y} = L_{F,Y}[Z \mapsto L(Z)] \subseteq (A \cup C')^\sharp.$$

Then, let us substitute words of $L'_{F,Y}$ for the occurrences of each letter $(F, Y)$ in words of $L_X$, where $Y \in F$ and $F \in \mathcal{F}$. We have that

$$L(X) = L_X[(F, Y) \mapsto L'_{F,Y}] \subseteq A^\sharp.$$

Since for every $Y \in F \subseteq C$ and $F \in \mathcal{F}$, the grammar $G_{F,Y}$ is linear, each $L_{F,Y}$ is a BCFL by Lemma 3.1. By the induction hypothesis, each $L(Z)$ with $Z \in N'$ is also a BCFL. Finally, $L_X$ is a BCFL since every ordinary CFL is a BCFL. Since BCFLs are closed under substitution, cf. [22], it follows now that $L(X)$ is a BCFL. $\square$

COROLLARY 4.2 *A language of scattered words is a BCFL iff it can be generated by an MCFG having no reproductive nonterminals.*

*Proof.* The sufficiency of the claim is immediate from Theorem 4.1. In order to prove the necessity, recall from [22] that every BCFL of scattered words over a finite alphabet $A$ can be generated by a BCFG $G = (N, A, R, S, F)$ such that for every strong component $C$ containing a nonterminal in $F$ and every rule $X \to p$ with $X \in C$, $p$ contains at most one occurrence of a nonterminal in $C$. Then let $G' = (N, A, R, S, \mathcal{F})$, where $\mathcal{F}$ is the set of all subsets of $N$ containing at least one nonterminal in $F$. Clearly, $G'$ is an MCFG equivalent to $G$ containing no reproductive nonterminal. $\square$

COROLLARY 4.3 *An MCFL $L$ of scattered words is a BCFL iff $L$ is bounded. A bounded language of scattered words is a BCFL iff it is an MCFL.*

*Proof.* Suppose that $L$ is a bounded MCFL of scattered words. It is known, cf. [23], that $L$ can be generated by a non-reproductive MCFG. Thus, $L$ is a BCFL by Theorem 4.1.

Suppose now that $L$ is a BCFL of scattered words. Then, as shown in [22], $L$ is bounded. $\square$

The previous corollary answers a question in [23].

## 5  Operational characterization of BCFLs of scattered words

In this section, we provide a Kleene-type characterization of the class of BCFLs consisting of well-ordered, or scattered words. We show how these languages may be constructed using ordinary context-free languages and the $\omega$-power operation defined earlier. We will also define expressions denoting BCFLs of well-ordered and scattered words.

LEMMA 5.1 *A language $L \subseteq A^\sharp$ is a BCFL of scattered words iff it can be constructed from the singleton languages $\{a\}$ for $a \in A$ by substitution into ordinary CFLs and linear BCFLs: Suppose that $B$ is a finite alphabet and $L_0$ is a CFL or a linear BCFL over $B$, and suppose that for each $b \in B$, $L_b \subseteq A^\sharp$ – then construct the language $L_0[b \mapsto L_b] \subseteq A^\sharp$.*

*Proof.* The sufficiency follows from the fact that the languages $\{a\}$, for $a \in A$ are BCFLs of scattered words as is every ordinary CFL or linear BCFL, and that both the class of all BCFLs and the class of all languages of scattered words are closed under substitution.

In order to prove the necessity, without loss of generality we may assume that $L \subseteq A^\sharp$ is a nonempty BCFL of scattered words which does not contain $\epsilon$. Thus, by a result in [22], $L = L(G)$ for a BCFG $G = (N, A, R, S, F)$ having no useless nonterminals and such that $L(X) = \{u \in A^\sharp : X \Rightarrow^\infty u\}$ contains at least one nonempty word for each $X \in N$. Since $G$ contains no useless nonterminals, it follows that each $L(X)$ consists of scattered words, and well-ordered words if every word in $L$ is well-ordered. Indeed, if $L(X)$ contains a quasi-dense word $u$, then since there exist words $v, w$ with $S \Rightarrow^\infty vXw$, $L$ contains the quasi-dense word $vuw$, contrary to our assumptions. Moreover, if $u$ is not well-ordered, then $vuw$ is also not well-ordered.

Let $X \in N$ be a nonterminal of height $n$. We show by induction on $n$ that the language $L(X) = \{u \in A^\sharp : X \Rightarrow^\infty u\}$ can be constructed as claimed. Suppose that we have proved this claim for all nonterminals of height $< n$. Let $C$ denote the strong component of $X$. If $C$ contains a nonterminal in $F$, then for any $Y \to p$ in $R$ with $Y \in C$, the word $p$ contains at most one occurrence of a nonterminal in $C$, cf. [22]. Then let $N'$ denote the set of all nonterminals $Y$ not in $C$ accessible from $X$, and consider the BCFG $G' = (C, N' \cup A, R', X, F')$, where $R'$ is the set of all rules in $R$ whose left side is in $C$ and $F' = C \cap F$. This grammar is linear so that $L' = L(G')$ is a linear BCFL. It is clear that $L(X)$ can be constructed from $L'$ by substituting the language $L(Y)$ for all $Y \in N'$ (and the language $\{a\}$ for all $a \in A$).

Suppose now that $C$ contains no nonterminal in $F$. Then consider the ordinary CFG $G' = (C, N' \cup A, R', X)$ where $N'$ and $R'$ are defined as above. Now $L(X)$ is the language obtained by substituting $L(Y)$ for each $Y \in N'$ in $L(G')$. $\square$

LEMMA 5.2 *A language $L \subseteq A^\sharp$ is a BCFL $L$ of well-ordered words iff it can be constructed from the singleton languages $\{a\}$ for $a \in A$ by substitution into ordinary CFLs and linear BCFLs consisting of well-ordered words.*

*Proof.* The proof of the sufficiency is the same as above. In order to prove the necessity, we argue as before using the fact that if $C$ contains a nonterminal in $F$, then for any $Y \to p$ in $R$ with $Y \in C$, the word $p$ contains at most one occurrence of a nonterminal in $C$, and it it occurs, it is the last letter of $p$, cf. [22]. $\square$

THEOREM 5.3 *A language $L \subseteq A^\sharp$ is a BCFL of well-ordered words iff it can be generated from the languages $\{a\}$, $a \in A$ by substitution into ordinary CFLs and the operation of $\omega$-power.*

*Proof.* If $L$ can be generated from the languages $\{a\}$, $a \in A$ by applying the two operations $n$ times, then since the $\omega$-power operation can be described as substitution into the linear BCFL $\{b^\omega\}$ over the one letter alphabet $\{b\}$, it follows by Lemma 5.1 that $L$ is a BCFL. On the other hand, it is clear that all words in $L$ are well-ordered.

In order to prove the opposite direction, by Lemma 5.2, all we need to show is that substitution into a linear BCFL consisting of well-ordered words can be expressed by substitution into ordinary CFLs and the $\omega$-power operation. But this is clear from Corollary 3.4 (or Corollary 3.5). □

Expressions denoting ordinary CFLs, similar to the regular expressions denoting regular tree languages [25], were introduced in [27]. A variant of these expressions are the well-known *$\mu$-expressions* used in several branches of computer science including process algebra and programming logics. By adding the operation of $\omega$-power to $\mu$-expressions in an appropriate way, we now define expressions denoting BCFLs of scattered words.

Let us fix a countably infinite set of variables. For a finite alphabet $A$, let $\mu\omega T$ denote the set of all expressions generated by the grammar

$$T ::= a \,|\, \epsilon \,|\, x \,|\, T + T \,|\, T \cdot T \,|\, \mu x.T \,|\, T_0^\omega$$

where $a$ is any letter in $A$, $x$ ranges over the variables, and the $\omega$-power operation is restricted to *closed terms* $T_0$. (A term is closed if each occurrence of a variable $x$ is in the scope of a prefix $\mu x$.) The semantics of expressions is defined by induction in the expected way. When the free variables of an expression $t$ form the set $V$, then $t$ denotes a language $|t| \subseteq (A \cup V)^\sharp$. (We assume that $A$ is disjoint from the variables.) The prefix $\mu$ corresponds to taking least fixed-points: for an expression $t$ with free variables in $V$ and a variable $x$, $\mu x.t$ denotes the least language $L \subseteq (A \cup (V \setminus \{x\}))^\sharp$ such that

$$|t|[x \mapsto L] = L.$$

This language exists by the well-known Knaster-Tarski theorem, since the function mapping a language $L' \subseteq (A \cup (V \setminus \{x\}))^\sharp$ to $|t|[x \mapsto L']$ is monotone. (Here, we understand that when $x$ does not occur free in $t$, then $|t|[x \mapsto L']$ is just $|t|$. We do not need a symbol denoting the empty language since it is denoted by the expression $\mu x.x$, where $x$ is a variable. Also, note that when $|t| = L$ and $x$ is a variable that does not appear in $t$, then $|\mu x.(tx + \epsilon)| = L^*$, the union of all finite powers of $L$.)

Let $\mu T$ denote the fragment of $\mu\omega T$ obtained by removing the $\omega$-power operation. Clearly, every expression $t \in \mu T$ denotes a language of finite words. It is known that a language $L \subseteq A^*$ is a CFL iff there is some closed $t \in \mu T$ over $A$ with $|t| = L$ (see [27] for a closely related result). Using this fact together with Theorem 5.3, we immediately have:

COROLLARY 5.4 *A language $L \subseteq A^\sharp$ is a BCFL of well-ordered words iff there is a closed expression $t \in \mu\omega T$ over $A$ with $|t| = L$.*

We give some examples to illustrate Corollary 5.4. Suppose that the alphabet contains the letters $a, b, c$. Consider the following expressions:

$$\begin{aligned} t_0 &= \mu x.(a^\omega x b^\omega + \epsilon) \\ t_1 &= (\mu x.(a^\omega x b^\omega + \epsilon))^\omega \\ t_2 &= \mu y.(\mu x.(a^\omega x b^\omega + \epsilon) y c + \epsilon) \end{aligned}$$

They denote the languages

$$\begin{aligned} L_0 &= \{(a^\omega)^n (b^\omega)^n : n \geq 0\} \\ L_1 &= L_0^\omega = \{(a^\omega)^{n_0}(b^\omega)^{n_0}(a^\omega)^{n_1}(b^\omega)^{n_1} \ldots : n_i \geq 0\} \\ L_2 &= \bigcup (L_0^n c^n : n \geq 0\}. \end{aligned}$$

We now turn to BCFLs of scattered words.

THEOREM 5.5 *Suppose that $L \subseteq A^\sharp$. Then $A$ is a BCFL of scattered words iff $L$ can be generated from the languages $\{a\}$, for $a \in A$ by the following operations:*

1. *Substitution into ordinary context-free languages.*

2. *The operation $L \times L' = \{(u, v) : u \in L, v \in L'\}$, where $L, L' \subseteq A^\sharp$.*

3. *The operations $U \cup V$, $U \cdot V$ and $U^*$, where $U, V \subseteq A^\sharp \times A^\sharp$.*

4. *The operation $U^\omega$, where $U \subseteq A^\sharp \times A^\sharp$.*

*Proof.* For $U \subseteq A^\sharp \times A^\sharp$, let us denote by $\widehat{U}$ the language $\{u\#v : (u,v) \in U\} \subseteq (A \cup \{\#\})^\sharp$, where $\#$ is a new symbol. In order to prove that every language generated from the languages $\{a\}$, $a \in A$ by the above operations is a BCFL of scattered words, by Lemma 5.1 it suffices to show the following claims.

*Claim 1* If $L$ and $L'$ are BCFLs of scattered words and $U = L \times L'$, then $\widehat{U}$ is a BCFL of scattered words.

*Claim 2* If $U, V \subseteq A^\sharp \times A^\sharp$ are such that $\widehat{U}$ and $\widehat{V}$ are BCFLs of scattered words, and if $W = U \cup V$, $W = UV$ or $W = U^*$, then $\widehat{W}$ is also a BCFL of scattered words.

*Claim 3* If $\widehat{U}$ with $U \subseteq A^\sharp \times A^\sharp$ is a BCFL of scattered words, then $U^\omega$ is a BCFL of scattered words.

Regarding the first claim, note that if $G = (N, A, R, S, F)$ and $G' = (N', A, R', S', F')$ are BCFLs generating $L$ and $L'$, respectively, where $N$ and $N'$ are disjoint, and if $U = L \times L'$, then the BCFG $G'' = (N \cup N' \cup \{S_0\}, A \cup \{\#\}, R \cup R' \cup \{S_0 \to S\#S'\}, F \cup F')$, where $S_0$ is a new symbol, generates $\widehat{U}$. Moreover, each word in $\widehat{U}$ is scattered, since each is a concatenation of three scattered words.

As for the second claim, let $U, V \subseteq A^\sharp \times A^\sharp$ such that $\widehat{U}$ and $\widehat{V}$ are BCFLs of scattered words. If $W = U \cup V$ then $\widehat{W} = \widehat{U} \cup \widehat{V}$ and it is clear that $\widehat{W}$ is a BCFL of scattered words, since BCFLs of scattered words are closed under union. Now let $W = UV$. Since $\widehat{W}$ can be obtained from $\widehat{U}$ by substituting $\widehat{V}$ for $\#$, and since both BCFLs and the class of languages containing scattered words are closed under substitution, it follows that $\widehat{W}$ is a BCFL of scattered words. Finally, suppose that $W = U^*$ and let $G = (N, A \cup \{\#\}, R, S, F)$ be a BCFG for $\widehat{U}$. We may assume that the right side of each rule contains at most one occurrence of $\#$. Then let $S_0$ be a new nonterminal and define $G' = (N, A \cup \{\#\}, R', S_0, F)$, where $R'$ contains the rule $S_0 \to \#$, the rules $X \to p$ in $R$ such that $\#$ does not appear in $p$, and all rules $X \to pS_0q$ such that $X \to p\#q$ is in $R$. Then $L(G') = \widehat{W}$. Since every word in $\widehat{W}$ is a concatenation of a finite number of scattered words, $\widehat{W}$ contains only scattered words.

In order to prove the last claim, suppose that $U \subseteq A^\sharp \times A^\sharp$ and $\widehat{U}$ is a BCFL of scattered words. It is clear that each word in $U^\omega$ is scattered, since the underlying linear order of each word in $U^\omega$ is a scattered sum of scattered linear orderings. To show $U^\omega$ is a BCFL, let $G = (N, A \cup \{\#\}, R, S, F)$ be a BCFG generating $\widehat{U}$. Let $S_0$ be a new symbol and consider the BCFG $G' = (N \cup \{S_0\}, A \cup \{\#\}, R', S_0, F \cup \{S_0\})$, where $R'$ consists of the rule $S_0 \to S$, all rules $X \to p$ in $R$ such that $p$ does not contain $\#$, and all rules $X \to pS_0q$ such that $X \to p\#q$ is in $R$. (Without loss of generality we may assume that the right side of each rule in $R$ contains at most one occurrence of $\#$.) Then $L(G') = U^\omega$.

We have thus proved that every language that can be constructed from the primitive languages $\{a\}$, $a \in A$ by the operations given in the Theorem is a BCFL of scattered words. The fact that every BCFL of scattered words over $A$ can be constructed follows from Lemma 5.1 and Theorem 3.2. □

We may now introduce expressions denoting BCFLs of scattered words. In our definition, we also use expressions denoting sets of pairs of words. The expressions in $\mu\omega T'$ over the alphabet $A$ are defined by the following grammar:

$$\begin{aligned} T &::= a \mid \epsilon \mid x \mid T + T \mid T \cdot T \mid \mu x.T \mid P^\omega \\ P &::= T_0 \times T_0 \mid P + P \mid P \cdot P \mid P^* \end{aligned}$$

Here, $T_0$ stands for an expression of syntactic category $T$ without free variables. Expressions corresponding to the syntactic category $P$ denote sets of pairs of words. The semantics of the expressions should be clear. When $t \in \mu\omega T'$, we write $|t|$ for the language denoted by $t$.

COROLLARY 5.6 *A language $L \subseteq A^\sharp$ is a BCFL of scattered words iff there is a closed expression $t \in \mu\omega T'$ (of syntactic category $T$) over $A$ with $|t| = L$.*

We again give some examples. But first, let us introduce some abbreviations. When $t$ is an expression of syntactic category $T$, then let us define $t^\omega$ and $t^{-\omega}$ as the expressions

$(t \times \epsilon)^\omega$ and $(\epsilon \times t')^\omega$. Now let

$$\begin{aligned} t_0 &= (a^\omega \times b^{-\omega})^\omega \\ t_1 &= ((a \times b)^*(b \times a))^\omega \\ t_2 &= \mu x.(a^\omega x b^{-\omega} + \epsilon). \end{aligned}$$

The languages denoted by these expressions are:

$$\begin{aligned} L_0 &= \{(a^\omega)^\omega (b^{-\omega})^{-\omega}\} \\ L_1 &= \{(a^{n_0} b a^{n_1} b \ldots) \cdot (\ldots a b^{n_1} a b^{n_0}) : n_i \geq 0\} \\ L_2 &= \{(a^\omega)^n (b^{-\omega})^n : n \geq 0\}. \end{aligned}$$

## 6 Applications

Suppose that $L \subseteq A^\sharp$ is a language of scattered words. Then let $\mathrm{rmax}(L) = \sup\{\mathrm{r}(u) : u \in L\}$, so that $\mathrm{rmax}(L)$ is an ordinal at most $\omega_1$, the first uncountable ordinal. When $L = \emptyset$, we define $\mathrm{rmax}(L) = -\infty$. (We understand that $-\infty < \alpha$ and $-\infty + \alpha = -\infty$ for all ordinals $\alpha$.) In [22], it was shown that $\mathrm{rmax}(L)$ is finite for every nonempty BCFL of scattered words. We give a new proof of this result.

THEOREM 6.1 *Suppose that $L \subseteq A^\sharp$ is a scattered BCFL. Then $\mathrm{rmax}(L)$ is either finite or $-\infty$.*

*Proof.* We use Lemma 5.1. When $L = \{a\}$, for some letter $a \in A$, then $\mathrm{rmax}(L) = 0$. Suppose now that $L = K[b \mapsto L_b]$, where $K$ is a CFL or a linear BCFL over some alphabet $B$ and each $L_b \subseteq A^\sharp$ is a scattered BCFL. Suppose that we have already shown that $\mathrm{rmax}(L_b)$ is finite for each $b \in B$. If $K = \emptyset$ then $L = \emptyset$ and $\mathrm{rmax}(L) = -\infty$. If $K = \{\epsilon\}$ then $L = \{\epsilon\}$ and $\mathrm{rmax}(L) = 0$. If $K \not\subseteq \{\epsilon\}$ then let $B_0$ denote the set of all letters of $B$ that occur in some word $u$ of $K$ such that $L_b \neq \emptyset$ for all $b$ occurring in $u$. We have $\mathrm{rmax}(L) \leq 1 + \max\{\mathrm{rmax}(L_b) : b \in B_0\}$. □

For a scattered language $L \subseteq A^\sharp$, let us define $\mathrm{rrange}(L) = \{\mathrm{r}(u) : u \in L\}$.

THEOREM 6.2 *Suppose that $L \subseteq A^\sharp$ is a scattered BCFL. Then $\mathrm{rrange}(L)$ is a finite set of integers that can be computed from a BCFG generating $L$.*

*Proof.* We already know that $\mathrm{rrange}(L)$ is a finite subset of the integers. We use Lemma 5.1 to show it is computable. When $L = \{a\}$, for some letter $a \in A$, then $\mathrm{rrange}(L) = \{0\}$. Suppose now that $L = K[b \mapsto L_b]$, where $K$ is a CFL or a linear BCFL over an alphabet $B$, and each $L_b$ is a BCFL. Suppose that we have already computed the sets $\mathrm{rrange}(L_b)$, $b \in B$. Since both CFLs and linear BCFLs are effectively closed under substitution of finite languages of finite words, we may assume that none of the $L_b$ is empty, and none of them contains $\epsilon$.

Given a grammar for $K$ and grammars for the $L_b$, we may compute the set $\Gamma \subseteq P(B) \times P(B)$ consisting of all pairs $(H_0, H_1)$ such that for some word $u \in K$, $H_0$ is the set of all letters that occur a finite number of times in $u$, and $H_1$ is the set of all letters that occur an infinite number of times in $u$.

Suppose that $(H_0, H_1) \in \Gamma$, say $H_0 = \{b_1, \ldots, b_k\}$, $H_1 = \{c_1, \ldots, c_\ell\}$. Then for all families $n_i \in \text{rrange}(L_{b_i})$ and $m_j \in \text{rrange}(L_{c_j})$, $i = 1, \ldots, k$, $j = 1, \ldots, \ell$, take the integer $n = \max\{n_i, m_j + 1 : i = 1, \ldots, k, \; j = 1, \ldots, \ell\}$. Then $\text{rrange}(L)$ is the set of all these integers $n$, for all possible choices of $(H_0, H_1)$ in $\Gamma$, and for all possible choices of the $n_i$ and $m_j$.

The fact that $\text{rrange}(L)$ can be computed from a BCFG for $L$ now follows by the constructive proof of Lemma 5.1. □

Suppose that $L \subseteq A^\sharp$ is a scattered language. Then let $\text{rmin}(L) = \min\{\text{r}(u) : u \in L\}$, so that $\text{rmin}(L)$ is a countable ordinal. When $L = \emptyset$, we define $\text{rmin}(L) = \infty$. As a corollary to the previous result, it is clear that for a BCFL $L$ generated by a BCFG $G$, $\text{rmin}(L)$ is either finite or $\infty$, and that $\text{rmin}(L)$ can be effectively computed along the lines of the previous proof. However, for each $(H_0, H_1)$ in $\Gamma$, it suffices to compute now only one integer, $\max\{n_i, m_j + 1 : i = 1, \ldots, k, \; j = 1, \ldots, \ell\}$, where using the above notation, $n_i = \text{rmin}(L_{b_i})$ and $m_j = \text{rmin}(L_{c_j})$ for all $i$ and $j$. Then $\text{rmin}(L)$ is the minimum of all these integers.

**Acknowledgement** The first author would like to thank the hospitality of the University of Aizu during his visit when the research reported here and the writing of this paper took place.